# Reply to Comment on "The origin of bursts and heavy tails in human dynamics"[1]

Understanding human dynamics is of major scientific and practical importance and can be increasingly addressed in a quantitative fashion thanks to electronic records capturing various human activity patterns. The authors of Ref. [1] revisit the datasets studied in Ref. [2], making four technical observations. Some of the observations of Ref. [1] are based on the authors' unfamiliarity with the details of the data collection process and have little relevance to the findings of Ref [2] and others are resolved in a quantitative fashion by other authors [3].

The observation that 5% of the emails are sent within ten seconds from each other is interesting, but the explanation provided by Ref [1] stems from the author's unfamiliarity with the dataset [4]: whenever a user sends an email to multiple recipients, they appear in the database as emails sent at the same time. Very long recipient lists are broken, however, into independent batches, appearing as separate emails sent seconds apart. Thus each time a user sends an email to a large recipient list, it will appear to display a significant short timeframe activity. Given the user base, offline activity is very unlikely to play a prominent role (J.P. Eckmann, private communication). Note, however, that a short time deviation from the power law is a common feature of systems characterized by heavy tailed distributions. Given, however, that the anomalous behavior of these systems is rooted in the large $\tau$ regime, the short time behavior is largely irrelevant, a fact widely known and discussed in the literature [5]. Thus, *independent of the origin of the short time behavior*, we fail to see its relevance to the results of Ref. [2], whose main finding is the existence of a fat *tail* in the interevent or the return time distributions, which refers to the intermediate and long time rather than the short time behavior.

The highly anomalous nature of a distribution with a power law tail with exponent α=1 is widely known and obvious. Its resolution has been provided by Ref [3], demonstrating that the exact solution of the model introduced in Ref. [2] has an exponential cutoff accompanying the power law.

We agree with Ref [1] that given the limited datasets, a lognormal distribution may appear to offer just as good fit for some users as a power law with α=1. The reason is that both the log-normal and the reported power law predict exactly the same leading behavior $\tau^{-1}$, differing only in the functional form of the exponential correction: for a log-normal distribution the correction has the form $e^{-(\ln \tau - M)^2 / 2S^2}$ while the model of Ref. [2] predicts $e^{-\tau/\tau_0}$, with $\tau_0 = \left( \ln \frac{2}{1+p} \right)^{-1}$ [3]. One could probably convincingly distinguish the two for very long datasets, but not for the email data of Ref [4]. Note also that Ref. [1] finds that a lognormal applies mainly to users with high short time activity, suggesting that the fit simply does a better job at capturing the short time activity than a power law. As discussed above, this short timescale is of little relevance to the tail's behavior. We would like to emphasize, however, that the main finding of Ref. [2] was that queuing processes of direct relevance to human dynamics *predict* that the distribution should have a fat tail with leading behavior $\tau^{-1}$. Given that Ref [1] fails to propose an alternative mechanism indicating that a lognormal distribution could also emerge in a model with relevance to the discussed human communication patterns, the observation that a lognormal distribution offers an equally good fit for some users is a mere exercise in statistics, one that has little hope to be conclusive until considerably longer records of email patterns become available.

---

[1] The comment [1], submitted to Nature by its authors, was rejected unanimously by three referees. Given, however, that the authors decided to subsequently place the comment on the ArXiv, we reluctantly follow suit.

The observation that a *p*-parameter-dependent fraction of tasks are executed immediately was discussed in the Supplementary Material of Ref [2] and Ref. [1] simply reiterates the results of Ref. [3]. The exact solution [3] goes well beyond the numerical results offered in [1], predicting the percentage of the tasks that have waiting time $\tau=1$ in function of the parameter *p*. The comment fails to propose a resolution, and offers numerical results that are of little novelty, given that they simply confirm the exact solution of Ref. [3].

In the model a *p*-dependent fraction of tasks are executed immediately ($\tau=1$ waiting time), and only the rest of the tasks follow a power law. Is this behavior realistic, or represents an artifact of the model? A superficial comparison with the empirical data would suggest that this is an artifact, as measurements shown in Fig. 1a do not provide evidence of tasks that are immediately executed. However, Fig 1a represents the interevent times, and not the waiting times, corresponding to the time a task waits on a user's priority list, which is what the model of Ref. [2] predicts. In the case when the waiting time can be directly measured, like in the email [2] or mail [6] based correspondence, there is some ambiguity to the real waiting time. Indeed, in the email data, for example, we have measured as waiting time the time difference between the arrival of an email, and the response sent to it. From an individual's or a priority queue's perspective this is not the real waiting time. Indeed, consider the situation when an email arrives at 9:00 am, and the recipient does not check her email until 11:56am, at which point she replies to the email immediately. From the perspective of her priority list the waiting time was less than a minute, as she replied as soon as she saw the email. In our dataset, however, the waiting time will be 3 hours and 56 minutes. The email dataset allows us, however, to get a much better approximation of the real waiting times than the observed interevent times [7]. Indeed, for an email $e_1$ received by user A we record the time $t_1$ it arrives, and then the time $t_2$ of the first email sent by user A to any other user *after* the arrival of the selected email. It will be this time from which we start measuring the waiting time for email $e_1$. Thus if user A replies to $e_1$ at time $t_3$, we consider that the email's waiting time $\tau_{real}=t_3-t_2$, instead of $t_3-t_1$ considered in Fig. 1a. The results, shown in Fig 1b, display the same power law scaling with $\alpha=1$ as we have seen in Fig. 1a, but in addition there is a prominent peak at $\tau_{real}=1$, corresponding to emails responded to immediately. This suggests that what we could have easily considered a model artifact in fact captures a measurable feature of email communications. Indeed, a high fraction of our emails is responded immediately, right after our first chance to read them, as predicted by the priority model introduced in Ref. [2].

In summary, the comment fails to provide evidence that would be of key importance to the overall message of Ref. [2], that queuing processes are responsible for the bursty activity patterns of humans. Most important, it fails to follow up on the logic of the comment, and answer the most important open question that would follow: Where would a lognormal distribution come from?

A.-L. Barabási, K.-I. Goh and A. Vazquez

                                                  Center for Cancer Systems Biology,
                                                  Dana Farber Cancer Institute,
                                                  Harvard University,
                                                  Boston MA 02115.

[1] Stauffer D.B. Dean Malmgren, R. Amaral, L. A. N., http://arxiv.org/physics/0510216.
[2] Barabási, A.-L. *Nature* **435**, 207–211 (2005).
[3] Vazquez, A. Exact results for the Barabasi model of human dynamics, *Phys. Rev. Lett.* (in press), http://arxiv.org/abs/physics/0506126.
[4] Eckmann, J-P., Moses, E., Sergi, D., *Proc. Nat. Acad. Sci. USA* **101**, 14333-14337 (2004).
[5] R. Albert and A.-L. Barabási, *Rev. Mod. Phys.* **74**, 47-97 (2002).
[6] J. G. Oliveira & A.-L. Barabási, Human Dynamics: The Correspondence Patterns of Darwin and Einstein., *Nature* **437,** 1251 (2005); http://arXiv.org/abs/physics/0511006.
[7] A. Vázquez, J. G. Oliveira, Z. Dezsö, K.-I. Goh, I. Kondor & A.-L. Barabási, Modeling bursts and heavy-tails in human dynamics, http://arXiv.org/abs/physics/0510117.

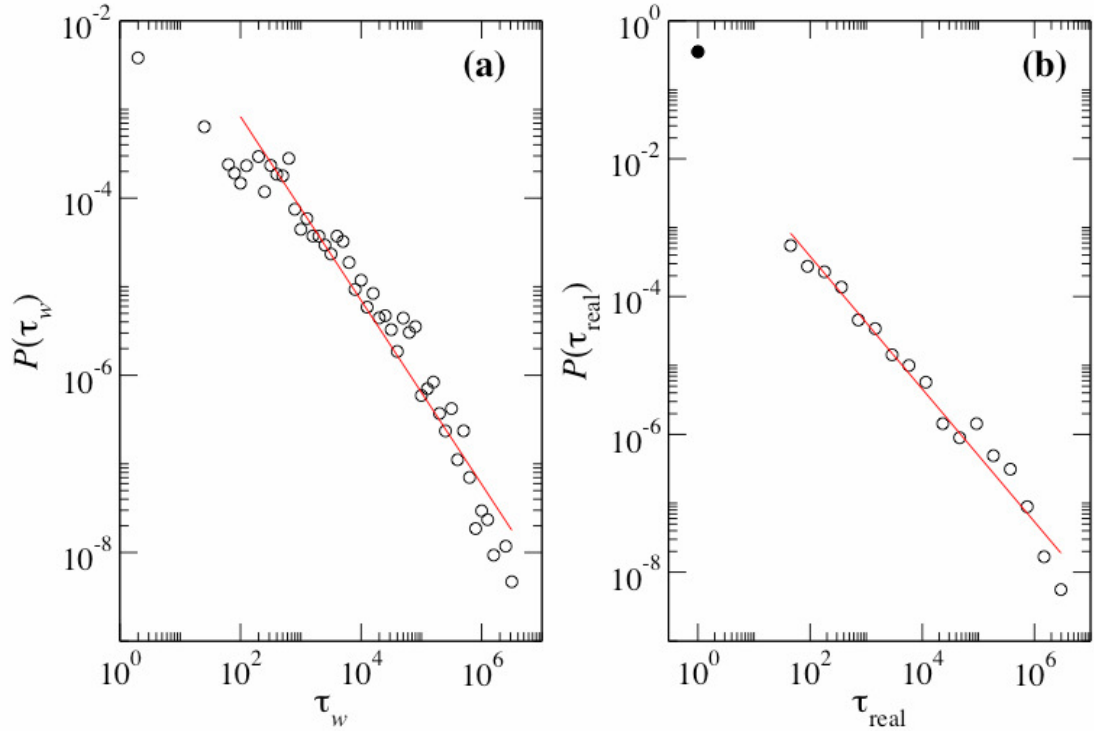

**Figure 1** Distribution of the response and arrival time intervals of an email user. (a) Given two email users A and B, the response times of user A to B are the time intervals between A receiving an email from B and A sending an email to B. The response time distribution of user A is then computed taking into account the response times to all users he/she communicates with. The continuous line is a power law fit with exponent $\alpha=1.0$. (b) The real waiting time distribution of an email in a user's priority list, where $\tau_{real}$ represents the time between the time the user first sees an email and the time she sends a reply to it. The black symbol shown in the upper left corner represents the messages that were replied to right after the user has noticed it.